\pdfoutput=1
\documentclass[sigconf,screen]{acmart}

\copyrightyear{2024}
\acmYear{2024}
\setcopyright{rightsretained}
\acmConference[IoT 2024]{14th International Conference on the Internet of Things}{November 19--22, 2024}{Oulu, Finland}
\acmBooktitle{14th International Conference on the Internet of Things (IoT 2024), November 19--22, 2024, Oulu, Finland}
\acmDOI{10.1145/3703790.3703797}
\acmISBN{979-8-4007-1285-2/24/11}

\usepackage[inline,shortlabels]{enumitem}
\usepackage[capitalise]{cleveref}
\usepackage{subcaption}
\usepackage{listings}
\usepackage{stfloats}
\usepackage{microtype}
\usepackage[utf8]{inputenc}

\begin{document}

\title{GoldFish: Serverless Actors with Short-Term Memory State for the Edge-Cloud Continuum}

\author{Cynthia Marcelino}
\orcid{0000-0003-1707-3014}
\affiliation{%
  \institution{Distributed Systems Group, TU Wien}
  \city{Vienna}
  \country{Austria}}
\email{c.marcelino@dsg.tuwien.ac.at}

\author{Jack Shahhoud}
\orcid{0009-0003-5857-2589}
\affiliation{%
  \institution{Distributed Systems Group, TU Wien}
  \city{Vienna}
  \country{Austria}}
\email{e1631081@student.tuwien.ac.at}

\author{Stefan Nastic}
\orcid{0000-0003-0410-6315}
\affiliation{%
  \institution{Distributed Systems Group, TU Wien}
  \city{Vienna}
  \country{Austria}}
\email{snastic@dsg.tuwien.ac.at}

\renewcommand{\shortauthors}{Marcelino, Shahhoud and Nastic}

\begin{abstract}
Serverless Computing is a computing paradigm that provides efficient infrastructure management and elastic scalability. Serverless functions scale up or down based on demand, which means that functions are not directly addressable and rely on platform-managed invocation. Serverless stateless nature requires functions to leverage external services, such as object storage and KVS, to exchange data. Serverless actors have emerged as a solution to these issues. However, the state-of-the-art serverless lifecycle and event-trigger invocation force actors to leverage remote services to manage their state and exchange data which impacts the performance, incurs additional cost and dependency on third-part services.
To address these issues, in this paper, we introduce a novel serverless lifecycle model that allows short-term stateful actors, enabling actors to maintain their state between executions. Additionally, we propose a novel serverless Invocation Model that enables serverless actors to influence the processing of future messages. 
We present GoldFish, a lightweight WebAssembly short-term stateful serverless actor platform which provides a novel serverless actor lifecycle and invocation model. GoldFish leverages WebAssembly to provide the actors with lightweight sandbox isolation, making them suitable for the Edge-Cloud Continuum, where computational resources are limited. Experimental results show that GoldFish optimizes the data exchange latency by up to 92\% and increases the throughput by up to 10x compared to OpenFaaS and Spin. 

\end{abstract}





\maketitle

\section{Introduction}\label{sec1}

Serverless Computing is a paradigm that offers automated infrastructure management, scale to zero, and elastic scaling. Typically, a Serverless application consists of a series of interconnected functions, also known as a Serverless Workflow, that exchange ephemeral data, which can be discarded after function processing. Due to the Serverless stateless design, functions in a workflow leverage external services such as object storage, message brokers, and Key-Value stores (KVS) to exchange ephemeral data and manage their state. Although external services provide benefits such as computing and IO separation, they add significant latency overhead~\cite{jonas2019cloud,WhereWeAreLiesAhead,scf,sonic}. Moreover, functions are not directly accessible; they are accessible via platform ingresses such as API Gateway and Load Balancer~\cite{Baresi,rise,onestep}, thus making direct communication more challenging. 
Serverless actors~\cite{akka,actor-base-designed,durable_functions,faasTrack,Ray,microactor,faasm} have emerged addressing these issues, thus enabling direct communication, state persistence, and concurrency management, which is crucial for Serverless functions. 

Actors~\cite{orig_actors,akkaActor} are isolated entities that can \textcircled{\small{1}} create other actors, \textcircled{\small{2}} directly communicate with other actors and \textcircled{\small{3}} influence the processing or state for the next received message~\cite{actor-base-designed,durable_functions,akkaActor,feasibilityActors,actorSurvey}. Serverless functions are \textcircled{\small{1}} stateless, \textcircled{\small{2}} non-addressable, and \textcircled{\small{3}} event-triggered~\cite{scf,Baresi,rise,onestep}. 

Existing Serverless actor approaches~\cite{akka,durable_functions,Ray,faasm,microactor} leverage the state-of-the-art Serverless design characteristics such as lifecycle~\cite{lambda_lifecycle} and event-trigger invocation~\cite{jonas2019cloud, rise,whatitis} to enable stateful and addressable actors. However, in the current Serverless function lifecycle~\cite{lambda_lifecycle}, Serverless functions are stateless. Therefore, existing actor-like Serverless approaches leverage remote services, incurring network overhead and costs with additional services.  

Existing approaches that enable persistent stateful functions include: 
\begin{enumerate*} [label=(\alph*)]
    \item \textit{Programming Models}~\cite{durable_functions, actor-base-designed, orleans, sehic2012ProgrammingModel} that abstract the function state handling from the developer and leverage external services to store it.
    Such Programming Models provide frameworks and libraries that automatically manage state persistence. While Programming Models simplify state management, they might introduce latency overhead as they rely on external services.
    \item \textit{Sidecars}~\cite{akka, Boki} systems that act as proxies and manage state interactions transparently, thus ensuring that state consistency and storage are handled outside the serverless function lifecycle, thereby reducing the function’s overhead. Despite their benefits, sidecars run alongside the function, consuming additional CPU and memory resources, which impacts the overall resource usage and might become a challenge at Edge-Cloud Continuum.
    \item \textit{Custom Sandboxes}~\cite{faasm,process-as-a-Service} ensure that functions can access and modify shared states in a controlled manner, providing isolation and, at the same time, enabling efficient state management. Although custom sandboxes might be lightweight, they are not interoperable with the current state-of-the-art platforms, limiting their usage on different serverless platforms such as Knative, OpenFaas, and OpenWhisk. 
%    \item Some \textit{Serverless Platforms}~\cite{microactor,cloudburstSF} integrate native support for stateful computing, allowing functions to maintain state across multiple invocations via dedicated or state-of-the-art storage mechanisms.
\end{enumerate*}
Although these approaches offer state preservation, allowing Serverless functions to execute as actors, they still rely on external services for state management, causing up to 95\% of the function latency~\cite{Faastlane,sand}. To fully utilize actor potential, the Serverless lifecycle must ensure actors can process multiple requests while preserving their state for a short period. Consequently, actors maintain states between executions, avoiding unnecessary state propagation of ephemeral data for interconnected events. 

%However, to enable actors to process multiple requests, actors must be addressable. 

%In the EdgeCloud Continuum, applications commonly leverage state propagation~\cite{CICCONETTI2022101689, RealizingStatefulEdge,microactor} approach where functions state is stored in nearby edge nodes and then propagated and replicated across the whole network. Nevertheless, they do not address actor reuse and specific messaging middleware tailored for actors and stateful functions.

%To address this, in this paper we propose a novel serverless actor lifecycle model that enables actor reuse and state management.

%In the Edge-Cloud Continuum, functions typically adopt a hybrid approach, combining direct communication and remote services. Functions use direct communication methods such as message queues for exchanging ephemeral data while relying on Cloud remote services such as object storage for long-term data storage. Thus, applications can effectively balance the limitations of the Edge computational resource constraints with the requirements for Serverless function communication~\cite{iot_middleware,lotus,RealTimeDataAnalytic}

%http://essca2018.servicelaboratory.ch/slides/essca18-slides-%C3%A1lvaroruizollobarren-scaled.pdf

%https://people.kth.se/~jspenger/slides/spenger23abs-slides.pdf

%
Existing Serverless actors enable direct communication and persistent statefulness by leveraging the existing Serverless lifecycle and event-triggered invocation. As a result, a series of interconnected events lead to multiple actor instances that still rely on external services to exchange ephemeral data and store their state, impacting the performance significantly.
To address these issues, in this paper, we propose novel Serverless lifecycle and invocation models that enable actors to process multiple interconnected requests and facilitate serverless actors to influence the processing of future messages.
Finally, we present GoldFish, a lightweight actor-based Serverless platform that executes serverless functions as actors. The main contributions of this paper include:

\begin{itemize}

   %short-term stateful support correlation 
    % 

    \item \textit{LCM:} \textit{A novel Serverless Lifecycle Model} that natively executes serverless functions as actors. It allows serverless actors to preserve a short-term state between the executions, thereby reducing multiple actor instantiating for multiple requests. 

    %Thus, enabling Serverless actors to preserve the state between executions instead of relying on external services. Thus, reducing latency and network overhead.

    \item SIM: \textit{A novel Serverless Invocation Model} that allows actors to influence the processing of future messages, enabling them to handle multiple messages. Busy actors can reject future messages or queue them for processing when available. Hence, SIM facilitates the processing of a set of connected events, such as in Serverless Workflows, and thus, it maintains the context and continuity of event processing.
    
    %native actor model specific communication in serverless computing
    \item \textit{GoldFish:} \textit{A WebAssembly Serverless Actor Platform} that leverages Wasm to provide a lightweight isolation. GoldFish architecture leverages the LCM model to enable serverless functions to execute as actors. 
    Furthermore, GoldFish introduces its dedicated message middleware that enables direct communication and leverages SIM to enable actors to influence the processing of future messages, thus processing multiple messages.
    
\end{itemize}

This paper has eight sections. 
\cref{motivation} presents the illustrative scenario and research questions.
\cref{archiecture} describes GoldFish Serverless lifecycle and invocation model as well as architecture overview. 
\cref{runtime} describes the lifecycle management and the event-triggered message invocation introduced by GoldFish and their usage. 
\cref{impl} shows the prototype implementation details.
\cref{evaluation} discusses the experiments and evaluation, 
\cref{related_work} presents related work. 
\cref{conclusion} concludes with a final discussion and future work.

\section{Motivation} \label{motivation}

\subsection{Illustrative Scenario}\label{scenario}

To better motivate our research, we present a use case for real-time field monitoring and disease detection in smart agriculture. To achieve this, IoT devices are strategically positioned throughout the fields to detect crop properties such as soil moisture, temperature, humidity, and sunlight. A Serverless workflow is employed to identify and respond to these agricultural needs.

Our workflow utilizes four Serverless functions, partially executed on the Edge close to the data source to reduce communication latency and partially executed on the Cloud. Edge tasks are responsible for processing large real-time data streams, sensor data, and simple disease detection. On the other hand, tasks that require more powerful computing resources, such as model training and inference, are carried out in the Cloud. Our motivating scenario is inspired by a Serverless Workflow for real-time environmental monitoring ~\cite{smartFarming}.

\begin{figure}[!tb]
\centering
\includegraphics[width=.7\linewidth]{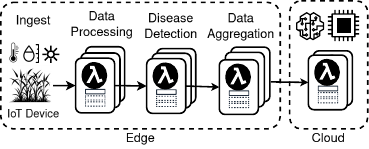}
\caption{Simplified Serverless Workflow for Disease Control for Smart Agriculture}
\label{fig:env_monitoring}
\end{figure}

In \cref{fig:env_monitoring}, in \textit{Ingest} stage, real-time data captured by IoT devices are transmitted to edge nodes via a streaming framework, where serverless functions responsible for \textit{Data Processing} are activated to execute tasks such as filtering, labeling and join the sensor data close to the source, thus reducing latency. Then \textit{Disease Detection} function processes part of the data, identifying specific patterns such as temperature and soil moisture. Then, \textit{Disease Detection} functions send data to \textit{Data Aggregation} functions, which combines current and historical data to enhance accuracy and reliability of the results. Finally, the processed data is transmitted to the cloud, where more resource-intensive tasks are performed, such as AI model inference to enhance disease analysis.

GoldFish decreases this workflow latency by enabling actors to process connected message events within a single actor. By softening some Serverless properties, such as statelessness, actors can keep a short-term state between executions, avoiding the need for remote services to exchange data. Additionally, actors can influence the processing of the next message, choosing to keep it in the queue for processing or reject it completely, allowing another actor to process the message. Thus, GoldFish enhances performance while maintaining the serverless nature of the function, as it still scales down to zero when not in use. On the other hand, it avoids the network overhead associated with external remote services for state persistence and data exchange. This tradeoff enhances performance without compromising the fundamental benefits of Serverless Computing.

\subsection{Research Challenges}
%The actor model's ability to deliver messages across distributed systems aligns with the dynamic and decentralized Serverless at Edge-Cloud Continuum~\cite{akkaActor}. However, the current Serverless design limits its full potential~\cite{virtual-actors,feasibilityActors,actorSurvey}. 
We identify the following research challenges to enable Serverless actors to maximize their performance in the Edge-Cloud Continuum.

\textit{RC-1: How to enable short-term stateful Serverless actors in the Edge-Cloud Continuum while preserving Serverless characteristics such as scale-to-zero?}

The current Serverless function lifecycle supports either succeeded or failed states, which leads to platforms creating multiple function instances for handling multiple function executions. 
Current approaches for Serverless actors preserve their state in remote storage and load the previous state in the new instance. Virtually, the new actor instance has the previous state, but physically, it is a new process on the host. Due to the current Serverless lifecycle design limitation, every request is a new actor, which requires actors to leverage external services to maintain their state~\cite{actor-base-designed,microactor,lambda_lifecycle}. Communication with external services causes the most function latency and additional costs, significantly affecting the performance and cost-efficiency of Serverless workflows. Relying on external services for state management adds latency, complexity, potential points of failure and costs due to frequent data retrieval~\cite{Faastlane,sand}.
Short-term stateful actors allow for state preservation within the actors themselves, eliminating the need for external services for state persistence. This minimizes the number of created instances, decreases latency and costs, and preserves resources at the edge. Hence, actors can scale to zero in the absence of invocations while still providing the advantages of stateful functions.  

\textit{RC-2: How can we enable direct communication between actors while allowing them to influence the processing of future messages?}

Direct communication among serverless actors requires addressability. By enabling direct message exchanges between actors, they avoid using external services to exchange data, thus reducing latency and network overhead. Nevertheless, the state-of-the-art event-triggered Serverless function invocation enables single message delivery, which means the platform cannot decide which function executes the message. To enable actors to influence future messages, the event-triggering middleware must \textcircled{\small{1}} forward to the actor for processing, \textcircled{\small{2}} enable actors to keep the message in the middleware until the actor becomes available again, or \textcircled{\small{3}} forward to another actor in case of rejection by the existing actor. By enabling actors to influence the processing of the message, users can decide to process connected message events in same actors, thus decreasing latency and network traffic overhead, crucial for enhancing performance in sensitive edge environments~\cite{scf,sand,Cwasi2023}.

\textit{RC-3: How to provide lightweight isolation while enabling the full potential of Serverless actors in the Edge-Cloud Continuum?} 

Isolation is critical to ensure that failures in one actor do not impact others. 
WebAssembly (Wasm) provides a secure sandboxed environment that reduces the overhead associated with traditional container-based isolation methods. Wasm lightweight isolation allows serverless actors to execute with reduced cold start, latency and resource consumption, which is crucial for the Edge-Cloud Continuum. Furthermore, actors can profit from the reduced cold starts Wasm, decreasing the actors startup time~\cite{Cwasi2023,PushingWebAssembly,,wasmCommonLayer}.

% TODO: It is still serverless in the sense if there are no invocation it scales to 0, in typical serverless you cannot control which function will process and you cant control how long the function will be alive. 
% one message vs sequence of message: i want to process all in one actor message. 
% support set of connected events

% 1. invocation model. 2. lifecycle 3. architecture
% temporarly stateful 

\section{GoldFish Serverless Models and Architecture Overview}\label{archiecture}

\subsection{GoldFish Serverless Lifecycle Model}

Golfish Serverless Lifecycle Model (LCM) provides an enhanced Serverless lifecycle specifically tailored for serverless actors to preserve their state between multiple executions while they are still alive. LCM still maintains Serverless characteristics such as elastic scaling and scale-to-zero while enabling actors short-term state memory. Thus, LCM optimizes resource usage, reduces latency, and improves performance and scalability across the dynamic environments of the Edge-Cloud Continuum by ensuring that existing actors are efficiently utilized and consequently minimizing the overhead associated with creating new actors.

\subsubsection{GoldFish Actor.} 
\begin{figure}[!tb]
\centering
\includegraphics[width=.8\linewidth]{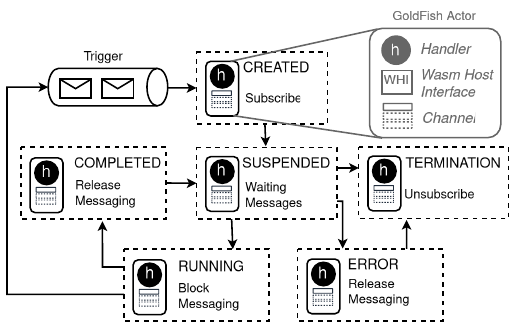}
\captionsetup{justification=centering}
\caption{GoldFish Serverless Lifecycle Model}
\label{fig:lifecycle}
\end{figure}

\cref{fig:lifecycle} shows GoldFish Actor and LCM Serverless Lifecycle. GoldFish actor is one entity composed of \textit{Channel}, \textit{Wasm Host Interface}, and \textit{Handler}.

\paragraph{Channel} It is identifiable by a unique ID and serves as a dedicated communication channel for the actor. It enables actors to carry their previous state to the next one. Proactive message blocking ensures that each actor processes only one message at a time, preventing data races and maintaining the integrity of the execution process.

\paragraph{Wasm Host Interface (WHI)} It is a sidecar process that creates the Wasm VM,
allowing for secure, isolated execution of the Wasm binary. It acts as a mediator between the Wasm binary and the channel, forwarding the input and output from the binary to the message channel. Upon receiving a message, WHI sends a signal to the Middleware to temporarily block any new incoming messages, ensuring actors process only a single message at a time.

\paragraph{Handler} It encapsulates the user-defined code compiled into a Wasm binary file. Functions execute in a Wasm sandbox, which means a controlled environment that limits access to the host system, receiving inputs and producing outputs through the \textit{Wasm Host Interface}.

\subsubsection{GoldFish Serverless Lifecycle Phases.} \cref{fig:lifecycle} shows LCM actor phases from the initialization to the termination. LCM phases are designed to ensure actor isolation and enable the (short-term) state management, key properties of actor model ~\cite{feasibilityActors,actor-base-designed,actorSurvey}. Each phase is responsible for specific tasks described below.
\paragraph{CREATED} This initial phase prepares the actor for operation and reserves the resources necessary. In addition, the actor receives a unique ID, which is later used for actor communication.

\paragraph{SUSPENDED} In this phase, the actor is not currently processing any tasks but is ready and waiting for new input or to terminate the actor if it remains in this phase for a long time. The period the actor remain in suspended phase is determine by the user. \emph{SUSPENDED} phase enables actors to keep the actor active but not running, it is waiting for incoming events to become active. This phase is essential for managing the efficient allocation of resources, enabling GoldFish to quickly respond to new messages without the overhead of the initialization phase. 

\paragraph{ERROR} In this phase, the actor has failed either during startup or execution. To enable message reprocessing, the actor releases the message and moves to \textit{TERMINATION} as a self-destroy mechanism. 
\paragraph{RUNNING} During this phase, GoldFish wakes up the actor from the \textit{SUSPENDED} phase and forwards the message to the actor. In this phase, it is where the actual data processing or task execution takes place. Additionally, in this phase, actors can create other actors by sending an addressed message to GoldFish Middleware.
\paragraph{COMPLETED} Once the messaging process is completed, the actor sends the results to the GoldFish middleware and signals its availability for further tasks. Then the actor can transition back to the SUSPENDED phase.
\paragraph{TERMINATION} The final phase of the lifecycle, where the actor stops receiving new messages, deregisters itself. In this phase, GoldFish releases resources and updates the actor state to reflect that the actor is no longer active.

\subsection{GoldFish Serverless Invocation Model}

The GoldFish Serverless Invocation Model (SIM) design ensures actors only handle one message at a time, which means concurrent requests is only possible with multiple actors, thereby avoiding concurrency issues and maintaining state integrity during the message delivery. Specifically, it enables processing multiple messages within a single actor message rather than handling each in isolation. Hence, SIM supports a set of connected events, facilitating more efficient workflow execution. Moreover, SIM enables serverless actors to influence future messages by keeping the message waiting to be executed, thus avoiding the use of remote services to store state and exchange data. As a result, it optimizes resource usage and reduces latency, which is crucial for improving the performance of functions in the Edge-Cloud Continuum.

\begin{figure}[!b]
\centering
\includegraphics[width=.6\linewidth]{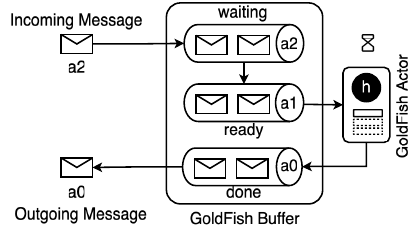}
\captionsetup{justification=centering}
\caption{GoldFish Serverless Invocation Model}
\label{fig:invocation}
\end{figure}
~\cref{fig:invocation} shows how SIM introduces a new way of triggering serverless actors in response to events such as incoming messages. The GoldFish SIM model ensures that new actors are created only when necessary while existing actors are reused by introducing a invocation Middleware  with three queues: \emph{waiting}, \emph{ready} and \emph{done}. GoldFish SIM model enables GoldFish Buffer to identify the availability and state of actors via the actor lifecycle phase. If the actor is \textit{SUSPENDED}, it transitions to the \textit{RUNNING} phase to handle the message. If the actor is busy, the Buffer keeps the message or forwards it to another available actor, ensuring seamless processing without message loss. Thus, SIM invocation model enables Serverless actors to process connected messages such as in a Serverless Workflow. To avoid a long waiting time, the Buffer has a time and message size limit defined by the user; once the time has reached, a new actor instance is created instead of reusing an existing actor.

\subsection{GoldFish Architecture Overview}

GoldFish leverages actor model properties such as addressability, isolation, and state to enhance serverless function execution by transforming them into Serverless actors~\cite{akkaActor,feasibilityActors,actorSurvey}. Each serverless actor in GoldFish is uniquely identifiable, allowing for direct, addressable communication, thereby facilitating efficient data and message exchanges across the actors in the Edge-Cloud Continuum.

\begin{figure}[!b]
\centering
\includegraphics[width=.5\linewidth]{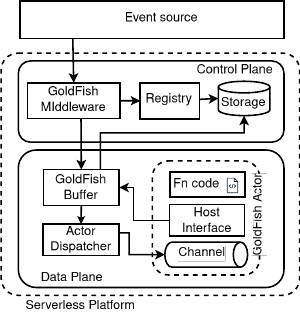}
\caption{GoldFish Architecture Overview}
\label{fig:architecture}
\end{figure}

The GoldFish architecture, shown in \cref{fig:architecture}, leverages Wasm to provide an isolated and secure sandbox for each actor. Moreover, GoldFish’s LCM manages the lifecycle of serverless actors from initialization to termination. GoldFish LCM enables Serverless actors to retain and efficiently manage their state, thus facilitating complex functions that require persistent state across sessions.

\subsubsection{GoldFish Components.} GoldFish is composed of main components: \textit{GoldFish Middleware, Registry, GoldFish Buffer and Actor Dispatcher}.

\paragraph{GoldFish Middleware} It accepts messages and ensures the messages are routed to the buffer in the correct node. When a GoldFish Buffer initiates, it registers itself in the GoldFish Middleware in the control plane. This registration enables the middleware proxy to route messages accurately to the designated actor dispatcher node.

\paragraph{Registry} It maintains a reference to the middleware across different nodes. When a middleware initiates, it registers itself within the registry. This registration enables the middleware proxy to route messages accurately to the designated actor dispatcher.

\paragraph{GoldFish Buffer} It is a queue for the busy actors, keeping waiting messages, thus allowing actors to influence the sequence of messages. When GoldFish Middleware receives a message, it passes it to the GoldFish Buffer if there is enough processing capacity. The Buffer then checks if the actors can handle new messages and sends them to the Actor Dispatcher. If the actors are unable to process new messages, the message is rejected. When a message is rejected, it is either kept waiting in the buffer or sent to another Actor Dispatcher until it is accepted. The message processing is defined by the actor, who can choose to receive the next message or reject it.

\paragraph{Actor Dispatcher} It manages the actors and their phases. The Actor Dispatcher receives the messaging events, identifies whether the actor exists by its unique ID, and forwards the message. The Actor Dispatcher updates Actor references in storage that are available via the control plane.

\section{GoldFish Mechanisms}\label{runtime}

GoldFish leverages the LCM Serverless Lifecycle Model and SIM Serverless Invocation Model to enable an actor native Serverless platform. GoldFish platform relies on two key mechanisms: LCM Serverless Lifecycle Phases Management and the  GoldFish SIM Serverless Event-triggered Message Invocation. 
%The GoldFish LCM Serverless Lifecycle Phases Management handles the lifecycle of serverless actors from the initialization to the termination. Meanwhile, the GoldFish SIM Serverless Event-triggered Message Invocation provides a mechanism for triggering serverless actors in response to events, enabling actor creation and direct communication between the actors. 

\subsection{GoldFish LCM Lifecycle Phases Management}

\begin{figure}[!b]
\centering
\includegraphics[width=.6\linewidth]{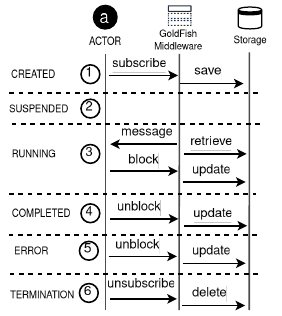}
\captionsetup{justification=centering}
\caption{GoldFish Serverless Lifecycle Management}
\label{fig:lifecycle_management}
\end{figure}

To execute Serverless functions as actors, GoldFish leverages the LCM to create and reuse actors.
\cref{fig:lifecycle_management} shows each phase and which services are necessary to enable the LCM. 

In \textcircled{\small{1}}, in \cref{fig:lifecycle_management},  when the actor is \textit{CREATED}, it subscribes to a specified channel with its unique ID. Then, GoldFish Middleware stores actor references for future usage. \textit{CREATED} is the initial phase where the platform executes tasks to prepare for the actor run, such as physical resource reservation and deployments. In the next phase in \textcircled{\small{2}}, the actor enters the \textit{SUSPENDED} phase, waiting for incoming messages for a period of time defined by the user. This is necessary to avoid actors to run constantly. In \textcircled{\small{3}}, a message is received, and the middleware retrieves information from the storage to identify the actor and forwards the message to the actor via the actor channel. Once the actor receives the message, it sends an event to the GoldFish Middleware to block new incoming messages. GoldFish middleware then updates the actor reference to the storage, finalizing this actor is busy and cannot receive any new message. In \textcircled{\small{4}}, the actor completes the message processing and sends a signal to GoldFish to unblock the actor. GoldFish Middleware updates the actor reference and removes the block. After this phase, the actor returns to phase \textcircled{\small{2}} to receive new messages. After a period defined by the user, the actor moves to the final phase \textit{TERMINATION} in \textcircled{\small{6}}. Phase \textcircled{\small{5}} represents an error state in the actor, the actor has either failed to startup or during execution. After entering the \textit{ERROR} phase, the actor unblocks the message in GoldFish Middleware which updates the actor reference in the storage.  In \textcircled{\small{6}}, the \textit{TERMINATION} phase, the actor unsubscribes to the channel. GoldFish Middleware deletes the specific channel and removes the actor reference from the storage. In this phase, the platform also releases reserved resources and removes any actor reference.

%\textit{Stateful Actors.} LCM is designed to enable direct communication, actor reuse, and actor recreation. GoldFish provides lifecycle management that natively enables functions to execute as actors. Additionally, GoldFish LCM enables platforms to leverage any existing mechanism to persist the actor state. In this paper, GoldFish leverages the current state-of-the-art Serverless state persistence as presented in ~\cite{cloudburstSF,durable_functions,akka,microactor}.

\subsection{GoldFish SIM Serverless Message Invocation}

SIM is a novel Serverless Invocation Model that enables Serverless actors to influence future messages. GoldFish Message Middleware leverages the SIM model to trigger and exchange messages between Serverless actors. GoldFish actors decide the processing of future messages based on the actor input; the GoldFish middleware decides whether to keep the message waiting in the buffer or forward it to the next actor.  

\begin{figure}[!b]
\centering
\includegraphics[width=.8\linewidth]{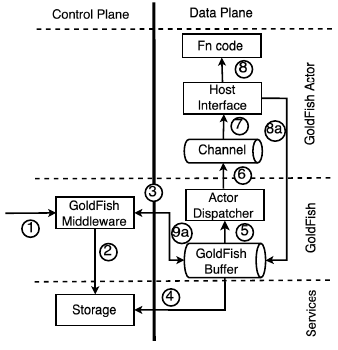}
\captionsetup{justification=centering}
\caption{GoldFish Distributed Messaging Middleware Flow}
\label{fig:messaging_flow}
\end{figure}

\cref{fig:messaging_flow} shows how GoldFish Middleware distributes the message from the event source to the user function code. In \textcircled{\small{1}}, an event arrives at the Middleware with the unique address of the actor. In \textcircled{\small{2}} the Middleware fetches from the storage existing actors information such as address and lifecycle phase to find out if any existing node contains such an actor already. In \textcircled{\small{3}}, the middleware forwards the message to either an existing actor that is suspended, an existing actor that signaled that they want to process it as the next message, or to the first free buffer that can potentially create a new actor to process such message. In 
\textcircled{\small{4}} the buffer queries the actor state to know whether it is immediately available if the actor wants to process the message next or reject it. In \textcircled{\small{5}} the buffer forwards the message to the Actor Dispatcher or keeps the message in memory for future processing. To avoid multiple storage queries, the buffer also forwards the actor information, which is necessary for the decision-making in the \textit{Actor Dispatcher}. In \textcircled{\small{6}}, the Actor Dispatcher creates an actor with its channel or forwards the message to an existing actor channel. This decision is made during the actor lifecycle phase. In \textcircled{\small{7}} the host interface receives the message from the channel and creates the Wasm VM. In \textcircled{\small{8a}}, if the actor wants to create another actor, e.g., send a message to another actor, the Wasm Host interface also communicates to the middleware to send a specific message. In \textcircled{\small{9a}}, the buffer forwards the message to the middleware, which starts the process for the new message receiving from in \textcircled{\small{1}}. In \textcircled{\small{8}} the \textit{Host Interface} starts the Wasm VM with the user function code.

\section{Prototype Implementation}\label{impl}

GoldFish is published as an open-source framework part of the Polaris SLO CLoud. Polaris itself is part of the Linux Foundation Centaurus project. GoldFish source code is available on GitHub\footnote{\url{https://github.com/polaris-slo-cloud/goldfish}}. 

The actor in GoldFish comprises a message channel, a Wasm host interface, and a Wasm binary containing the user function code. We utilize WasmEdge\cite{wasmedge} as the runtime, along with WasmEdge libraries, to create the Wasm VM. 
%The function code is compiled into a WebAssembly binary file (.wasm) and loaded into the Wasm VM via WasmEdge Runtime APIs. 
To ensure scalability, we use Docker\cite{docker} to run GoldFish actors, and the Rust wasmedge-sdk\cite{wasmedge_rust_sdk} facilitates interaction with WasmEdge. Events are sent to the middleware using WasmEdge Host Functions, which enable WebAssembly to call native Rust functions by passing them as imports to Wasm modules. The middleware is responsible for receiving and forwarding events to dispatchers, registering itself in the middleware registry upon startup. It communicates with Redis\cite{redis} to verify actor information such as phase and address and is implemented using GRPC interfaces. The middleware registry collects references to active middleware via GRPC and stores these references in Redis. 
%The middleware interface handles new requests received via GRPC, querying the registry to find the appropriate middleware and node for message delivery.
Actor dispatchers respond to events received by the middleware, creating an OCI Bundle with Docker that encapsulates the actor, ensuring interoperability with state-of-the-art platforms. Implemented in Rust, the dispatchers use Rust libraries to create GRPC interfaces that are available to the bus.

\section{Evaluation}\label{evaluation}

%\subsection{Overview}
We design our experiments to evaluate our GoldFish based on our illustrative scenario, shown in~\cref{scenario}, and on the most common invocation patterns of Serverless Computing: Sequential Executions and Fan-out execution, as discussed in~\cite{jonas2019cloud}. The goal of the evaluation is to measure the performance of the contributions LCM, SIM and Goldfish platform presented in \cref{sec1}.

\paragraph{Baselines \& Experimental Workflows} We compare GoldFish to OpenFaas\cite{openfaas} and Spin\cite{spin_v2}. We have chosen Openfaas to compare GoldFish with a standard container Serverless Platform that has wide support in the open-source community.  As GoldFish, Spin leverages WebAssembly, and therefore, it is important for GoldFish to compare with a framework that leverages similar technologies. We execute \emph{Chained Functions} and \emph{Serverless Workflow}, based on our illustrative scenario in~\cref{scenario}, for all three baselines (OpenFaaS, Spin and GoldFish) with three functions to simulate real-wold data-intensive Serverless use cases. In Chained Functions, we show the use case when a serverless functionA calls a serverless functionB. In Serverless Workflow, the next function is only executed once the previous function has finished. 
%In order to compare the performance using different message delivery frameworks, we use Redis Pub/Sub for delivering messages for OpenFaas and Spin. GoldFish middleware will be used in GoldFish.

\paragraph{Metrics} \emph{Latency} shows the execution time for the message passing between two actors. We use seconds and milliseconds for our latency experiments for Sequential and Parallel execution, respectively. Moreover,
\emph{Throughput} measures the number of executions a framework can process in a specific timeframe. We measure the performance of GoldFish under high load. The goal of Throughput experiments is to identify how many requests can the function process at a time and if there are bottlenecks in the proposed framework once the function load increases.

\subsection{Experiment Setup}
To evaluate GoldFish, we execute the designed experiments on a Ubuntu 22.04 LTS machine CPU ARM64~(AARCH64) with 8 GB of RAM, 4 cores, and 39 GB of storage.
The experimental functions and workflow are written in Rust for all the baselines. The baseline functions used for the evaluation expose REST API endpoints for receiving and processing requests from external sources. For the HTTP requests, we use Rust libraries for sending multiple parallel requests concurrently.
To ensure the consistency of the results and avoid bias, we executed the experiments seven times and calculated the average as the desired result.

\subsection{Experiment: Sequential Executions}\label{subsec:exp_seq}

In this experiment, we perform sequential request executions for our two experimental workflows: Chained Functions and Serverless Workflow.

\begin{figure}[ht]
\centering
\includegraphics[width=.8\linewidth]{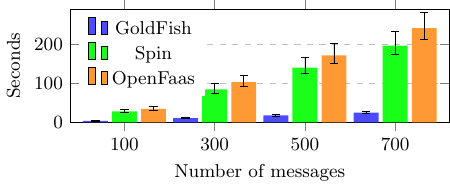}
\caption{Message Exchange Latency}
\label{fig:siwa_bus_message}
\end{figure}

\begin{figure}[!t]
\begin{subfigure}[t]{.23\textwidth}

\includegraphics[width=1.6\linewidth]{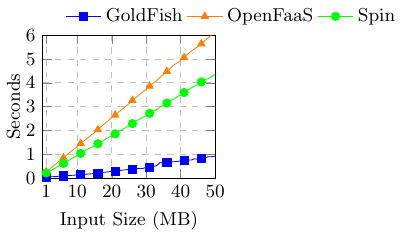}
\caption{Latency}
\label{fig:seq_chained_exec_latency}
\end{subfigure}
\hspace*{-1em}
\begin{subfigure}[t]{.23\textwidth}
\centering
\includegraphics[width=1\linewidth]{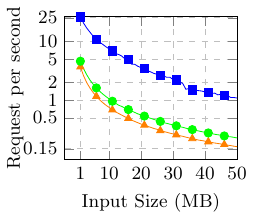}
\caption{Throughput}
\label{fig:seq_chained_exec_throughput} 
\end{subfigure}
\caption{Sequential Execution: Chained Functions}
\end{figure}

In ~\cref{fig:siwa_bus_message}, we show the latency of multiple sequential message requests processed by GoldFish Middleware, and the baseline Spin and OpenFaas. In the $x$ axis, we display the number of messages and, in $y$, the latency to process these messages. GoldFish Middleware shows latencies from approximately 3.56 to 25.04 seconds, while Spin displays an increase from about 28.06 to 196.25 seconds, and OpenFaas shows latency growing from roughly 34.58 to 241.00 seconds. The results show that GoldFish reduces latency up to 89\% when compared to the baseline.
\cref{fig:seq_chained_exec_latency} shows the input data size on the $x$ axis and the latency in seconds on the $y$ axis. GoldFish displays response times ranging from 0.039 to 0.919 seconds, OpenFaaS shows an increase from about 0.272 to 6.144 seconds, and Spin's response time grows from 0.218 to 4.362 seconds. The latency analysis reveals that GoldFish decreases the latency by up to 85\% and 79\% compared to OpenFaaS and Spin, respectively. 
These latency experiments show a significant latency reduction of GoldFish, with all three systems demonstrating a generally linear increase in response times, indicative of stable performance across the increasing load.
\cref{fig:seq_chained_exec_throughput} shows the throughput of GoldFish, OpenFaaS, and Spin as increasing the input size. The $x$ axis represents the input data size, while the $y$ axis shows requests per second. Over axis $x$, GoldFish's throughput decreases from about 25.93 to 1.09 requests per second, OpenFaaS declines from 3.68 to 0.16 requests per second, and Spin drops from 4.60 to 0.23 requests per second. All systems experience a linear decrease in throughput as the input size increases, indicating a linear throughput decrease with the input size. Additionally, GoldFish maintains a throughput up to 6.8 times higher than OpenFaaS and up to 4.7 times higher than Spin.

\begin{figure}[t]
\begin{subfigure}[t]{.23\textwidth}
\includegraphics[width=1.6\linewidth]{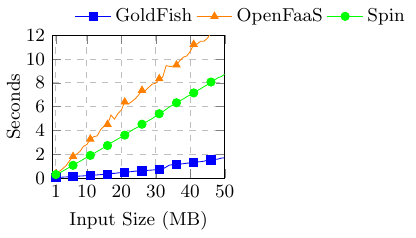}
\caption{Latency}
\label{fig:work_chained_exec_latency}
\end{subfigure}
\hspace{-1em}
\begin{subfigure}[t]{.23\textwidth}
\includegraphics[width=0.97\linewidth]{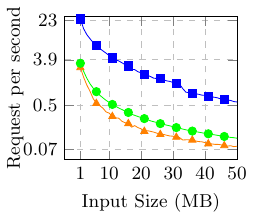}
\caption{Throughput}
\label{fig:workflow_chained_exec_throughput}
\end{subfigure}
\caption{Sequential Execution: Serverless Workflow}
\end{figure}

\cref{fig:work_chained_exec_latency} presents the input data size in megabytes on the $x$ axis and the response latency on the $y$ axis. As input size increases, GoldFish shows latency improvements ranging from 40 milliseconds to 1.71 seconds. OpenFaas displays latency from 363 milliseconds to approximately 12.5 seconds, while Spin maintains an increase from 299 milliseconds to 8.73 seconds. This experiment shows that GoldFish reduces latency by up to 86\% compared to OpenFaas and 80\% relative to Spin. 

\cref{fig:workflow_chained_exec_throughput} shows the throughput metrics, where the input data size is in megabytes on the $x$ axis and the requests per second on the $y$ axis. GoldFish displays a throughput decrease from 24.65 to 0.59 requests per second, while OpenFaas and Spin show reductions from 2.75 to 0.08 and from 3.34 to 0.11 requests per second, respectively. GoldFish presents up to 7.4 times higher throughput than OpenFaas and up to 5.4 times more than Spin.

\subsection{Experiment: Fan-out Parallel Executions}\label{subsec:exp_parallel}

In these experiments, we measure GoldFish scalability with fan-out parallel request executions for Chained Functions and Serverless Workflows. 

\begin{figure}[b]
\begin{subfigure}[t]{.23\textwidth}
\includegraphics[width=1.6\linewidth]{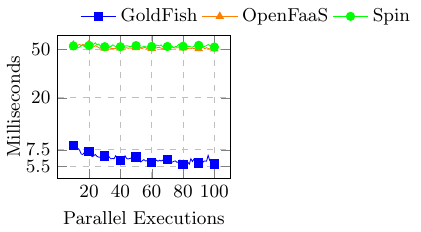}
\centering
\caption{Latency}
\label{fig:parallel_chained_exec_latency}
\end{subfigure}
\hspace{-1em}
\begin{subfigure}[t]{.23\textwidth}
\includegraphics[width=0.9\linewidth]{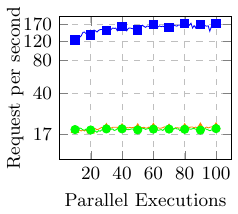}
\caption{Throughput}
\label{fig:parallel_chained_exec_throughput} 
\end{subfigure}
\caption{Parallel Execution: Chained Functions}
\end{figure}

\cref{fig:parallel_chained_exec_latency} presents the latency from the parallel execution experiments, where the $x$ axis represents the number of parallel executions and the $y$ axis reflects latency in milliseconds. \cref{fig:parallel_chained_exec_latency} that GoldFish maintains a relatively stable latency ranging from 6.9 milliseconds to around 5.75 milliseconds, even as the number of parallel executions increases. In comparison, OpenFaas and Spin exhibit slightly higher latency under higher loads, with OpenFaas and Sping showing a latency of around 50 milliseconds. GoldFish shows up to an 87\% reduction in latency compared to OpenFaas and Spin. 

In \cref{fig:parallel_chained_exec_throughput}, GoldFish maintains higher throughput, ranging from 123.45 to about 173.91 requests per second, which aligns with its efficient latency results under parallel operations in \cref{fig:parallel_chained_exec_latency}. OpenFaas and Spin also display consistent throughput, with OpenFaas and Spin presenting around 50 requests per second even when the function load increases in axis $x$. Overall, GoldFish has up to 9x higher throughput when compared to OpenFaas and Spin.

\begin{figure}[b]
\begin{subfigure}[t]{.23\textwidth}
\includegraphics[width=1.6\linewidth]{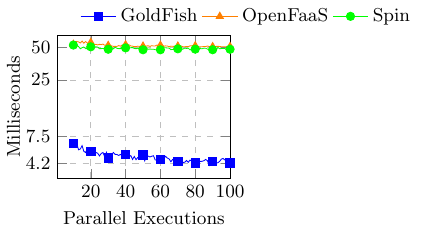}
\caption{Latency}
\label{fig:parallel_worfklow_exec_latency}
\end{subfigure}
\hspace{-1em}
\begin{subfigure}[t]{.23\textwidth}
\includegraphics[width=0.9\linewidth]{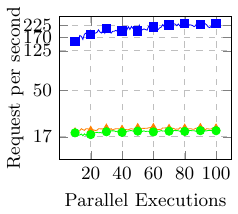}
\caption{Throughput}
\label{fig:parallel_worfklow_exec_throughput}
\end{subfigure}
\caption{Parallel Execution: Serverless Workflow}
\end{figure}

\cref{fig:parallel_worfklow_exec_latency} showcases the latency from parallel execution for Serverless Workflows, where the $x$ axis indicates the number of parallel executions and the $y$ axis measures the latency in milliseconds. GoldFish demonstrates stability in latency, which ranges from 6.4 ms to 4.23 ms as the parallel execution count increases.  In contrast, OpenFaas and Spin display higher latency similar to the nested functions, in \cref{fig:parallel_chained_exec_latency}, around 50ms. Compared to the baselines, GoldFish's latency is lower, showing an improvement of approximately 92\% for serverless workflows.

In \cref{fig:parallel_worfklow_exec_throughput}, GoldFish maintains a high throughput ranging from 156.25 to 236.41 requests per second. Both OpenFaas and Spin also show consistent throughput; however, they show around 20 requests per second, significantly lower than GoldFish. These results show that GoldFish has up to 10x higher throughput compared to the baselines, showing stability for high-load serverless workflows while maintaining high throughput and low latency.

\section{Related Work}\label{related_work}

%\subsection{Serverless Actor Model}
\paragraph{Serverless Actor Model}$\mu$Actor~\cite{microactor} introduces a lightweight stateful serverless platform able to execute actors not only on the cloud but also at the edge with limited resources such as microcontrollers. $\mu$Actor enables actors to send and receive messages from another actor via publish/subscribe mechanisms. Furthermore, actors may have access to additional devices such as sensors, actuators, databases, and DSP chips. Nevertheless, the introduced platform is not interoperable with the existing state-of-the-art platforms such as Knative, OpenFaas, and OpenWhisk, while Goldfish implements the actor a standard container which can be used by most of open source and comercial Serverless platforms.
Microsoft Azure's Durable Functions (DF)~\cite{durable_functions} introduces programming model abstractions to enable function state handling while ensuring reliable task progression. DF combines task and actor parallelism to create a fault-free function model. However, DF is specifically designed for the Azure platform, limiting its usage across other Serverless Platforms such as AWS Lambda, OpenFaaS, and OpenWhisk.
Akka~\cite{akka} introduces a side-car container that intercepts the incoming and outgoing traffic to manage the function state via external storage and proxies the traffic to the user function container. Nevertheless, Akka introduces an additional system that runs on an additional container, leading to potential increased resource usage, thus limiting its usage in the Edge-Cloud Continuum, where computational resources are limited. 
Ray~\cite{Ray} introduces a fully managed serverless platform tailored for AI that natively integrates the actor properties in the serverless functions, ensuring fault recovery and at-least-once message delivery mechanism. Ray preserves the state between the serverless AI workflow, wrapping multiple functions into one actor, such as extract and process frames, thus enabling low latency as functions are embedded in one actor.
%Nevertheless, this approach reduces the scalability of a single function, binding the scalability of the two functions together, which can lead to increased resource usage as all the embedded functions must be scaled together.
Although these approaches enable Serverless actors, they still rely on external services to persist the actor state even for ephemeral and intermediate data, thus increasing latency, costs, and digital waste. Goldfish keeps a short-term memory state in the actor so that actors can leverage the state to exchange ephemeral data exchange.

\paragraph{Stateful Serverless}
Faasm~\cite{faasm} introduces a stateful Serverless via faaslet and a two-tier state architecture for state and message exchange via faabric~\cite{faabric}. Faaslet provides lightweight isolation for each function, while the two-tier state architecture enables local and global function state storage based on the function location. Nevertheless, Faasm introduces customized isolation mechanisms incompatible with the OCI specs~\cite{oci-spec} of the current state-of-the-art serverless platforms. 
Cloudburst ~\cite{cloudburstSF} proposes a stateful  Serverless platform that leverages Anna~\cite{anna} Key-Value Store (KVS) for data exchange. Cloudburst replicates part of the cache locally for each function, allowing low-latency access, while remote data is accessed via Anna KVS. Although Cloudburst offers low latency and a highly scalable serverless platform, it might introduce duplicate cached data, leading to network overhead and duplicate serialization, a challenge for the limited resources of the Edge-Cloud Continuum. 
%DS2P~\cite{ds2p} proposes a decentralized stateful serverless platform that offloads serverless functions based on SLO-aware policies. DS2P estimates the function execution time on local and remote nodes, considering the additional latency required to retrieve the function state. However, DS2P focuses on scheduling mechanisms and leverages external services such as state-of-the-art KVS to enable data exchange and state persistence in the functions. 
Although the presented approaches enable stateful serverless, they focus on a persistent state, leading to network overhead, dependency on external systems, and additional costs. As Goldfish provides short-term state and multiple request executions, actors can keep their state for a short period between executions, avoiding the need for external service and thus improving performance significantly.

\section{Conclusion \& Future Work}\label{conclusion}

In this paper, we presented Goldfish, a short-term stateful Serverless for the Edge-Cloud Continuum that provides a novel Serverless Lifecycle Model (LCM) that allows actors keep a short-term state. Goldfish provides also SIM, a novel Serverless Invocation Model that enables actors to influence the processing of future messages, thus enabling one actor to process multiple requests. GoldFish leverages Wasm to provide a secure and isolated sandbox while enabling efficient ephemeral-data communication among serverless actors, thus optimizing performance and scalability in distributed environments.

Our evaluation demonstrates that GoldFish decreases latency and increases throughput, thereby enhancing performance in the Edge-Cloud Continuum. Specifically, GoldFish reduces latency by up to 92\% and increases throughput by up to 10 times.
GoldFish is specifically designed to address the requirements of the Edge-Cloud Continuum. 
Goldfish provides a lightweight Wasm sandbox, which fits the limited resource environment of the Edge Cloud Continuum. 

In the future, we plan to expand Goldfish into the 3D Edge Cloud Space Continuum. To achieve this, we intend to integrate Orbital Edge Computing (OEC) requirements, including satellite positioning, into the Goldfish platform requirements. This will enable Goldfish to execute workflows within the 3D Continuum seamlessly. Moreover, we intend to expand GoldFish by implementing a smart and serialization-free actor state. This enhancement will allow the platform to identify if the actors necessitate a remote state, thus preventing unnecessary state persistence. As a result, resource usage will be optimized and latency reduced by skipping the loading of actor states. Finally, we aim to integrate Goldfish into ML pipelines in Edge-Cloud Continuum to facilitate stateful data-intensive workloads such as~\cite{maresch2024vate}.

\begin{acks}
This work is partially funded by the Austrian Research Promotion Agency (FFG) under the project RapidREC (Project No. 903884).
This research received funding from the EU’s Horizon Europe
Research and Innovation Program under Grant Agreement No.
101070186. EU website for TEADAL: \url{https://teadal.eu}.
\end{acks}

\bibliographystyle{unsrtnat}
\bibliography{references}

%\printbibliography
\end{document}